# An investigation into the scientific landscape of the conversational and generative artificial intelligence, and human-chatbot interaction in education and research

Ikpe Justice Akpan, *Ph.D.*[1,*], Yawo M. Kobara, *Ph.D.*[2], Josiah Owolabi, *Ph.D.*[3], Asuama Akpam, *MA Ed*[4], and Onyebuchi Felix Offodile, *Ph.D.*[5]

1. Professor of Information Systems and Business Analytics, Kent State University, Ohio, USA; *(*Corresponding Author: 330, University Drive NE, New Philadelphia, OH 44663; Tel: +13303393391 EXT 47572; Fax: +13303393321; Email: iakpan@kent.edu).*
2. Assistant Professor of Operations Management and Statistics, Odette School of Business, University of Windsor, ON, Canada;
3. Associate Professor of Educational Evaluation, Faculty of Education, Open University of Nigeria, Abuja, Nigeria;
4. Research Associate, Ibom International Center for Research and Scholarship, Windsor, ON, Canada;
5. Professor of Operations Management and the immediate past Chair of the Department of Management & Information Systems, Kent State University, Kent, Ohio, USA.

**Abstract:**

Artificial intelligence (AI) as a disruptive technology is not new. However, its recent evolution, engineered by technological transformation, big data analytics, and quantum computing, produces conversational and generative AI (CGAI/GenAI) and human-like chatbots that disrupt conventional operations and methods in different fields. This study investigates the scientific landscape of CGAI and human-chatbot interaction/collaboration and evaluates use cases, benefits, challenges, and policy implications for multidisciplinary education and allied industry operations. The publications trend showed that just 4% (n=75) occurred during 2006-2018, while 2019-2023 experienced astronomical growth (n=1763 or 96%). The prominent use cases of CGAI (e.g., ChatGPT) for teaching, learning, and research activities occurred in computer science [multidisciplinary and AI] (32%), medical/healthcare (17%), engineering (7%), and business fields (6%). The intellectual structure shows strong collaboration among eminent multidisciplinary sources in business, Information Systems, and other areas. The thematic structure of SLP highlights prominent CGAI use cases, including improved user experience in human-computer interaction, computer programs/code generation, and systems creation. Widespread CGAI usefulness for teachers, researchers, and learners includes syllabi/course content generation, testing aids, and academic writing. The concerns about abuse and misuse (plagiarism, academic integrity, privacy violations) and issues about misinformation, danger of self-diagnoses, and patient privacy in medical/healthcare applications are prominent. Formulating strategies and policies to address potential CGAI challenges in teaching/learning and practice are priorities. Developing discipline-based automatic detection of GenAI contents to check abuse is proposed. In operational/Operations research areas, proper CGAI/GenAI integration with modeling and decision support systems requires further studies.

**Keywords**: generative artificial intelligence, quantum computing, disruptive technology, conversational chatbot, ChatGPT, human-robot interaction, big data analytics, technological transformation

## 1. Introduction

Artificial intelligence (AI) as disruptive technology is not new. However, what started in the mid-1900s through the early pioneers like Alan Turing (Breuel, 1990; Delipetrev et al., 2020) has evolved significantly over the years. The recent advances, engineered by technological transformation and statistical and quantum computing through conversational and generative AI (CGAI/GenAI) and human-like chatbots, such as the chat generative pre-trained transformer (ChatGPT) tends to disrupt the conventional approach in operations activities in diverse fields, including teaching, learning, and research (Li et al., 2023; Heng et al., 2023). Zhu et al. (2023) analyze the enormous computational advances and information theory that contribute to today's progress in AI development as a problem-solving tool and its application in education, business, and social lives through chatbots creation. Operational/Operations Research (OR) and decision sciences fields also have a long history of AI application to enhance informed decision-making (Al-Surmi et al., 2022; Pomerol, 1997). The pervasiveness of digital solutions and the advent of the fourth and fifth industrial revolutions brought integrated technologies, including the Internet of Things (IoT), virtual reality, and human-robot collaboration (Wienrich and Latoschik, 2021; Akpan and Offodile, 2024; Wan et al., 2017; Sheth et al., 2019). This advancement in AI and natural language processing (NLP) has enabled the development of AI agents, such as ChatGPT, which significantly enhances chatbot-human interaction capability, generating content across various domains, such as text, images, and enabling social actions (Heng et al., 2023).

Chatbots are computer programs that use AI models designed with the capability to generate different activities, including conversation with human users when prompted. These CGAI models are based on a larger framework, including machine learning (ML) and NLP (Kusal et al., 2022). NLP models can recognize speech, understand languages, and generate language and speech using computing algorithms. The algorithms enable systems to learn from data and improve performance over time as more data becomes available without being explicitly reprogrammed (Kang et al., 2020). This study explores the scientific literature landscape on conversational chatbots, GAI, and chatbot-human interaction in education and research: A multidisciplinary perspective and challenges of conversational and generative artificial intelligence (CGAI) in higher education using science mapping approach. Specifically, the study seeks to address the following five research objectives:

i. Evaluate the scientific literature publications (SLP) trend on CGAI application in educational services operations, including teaching and learning, and research/creative activities (RO1).
ii. Analyze the thematic structure of CGAI to highlight the use cases, benefits, and challenges and robot-human interaction in education and research (RO2).
iii. Determine the intellectual structure of research on CGAI in education (RO3).
iv. Formulate policy recommendations about the future of CGAI in education (RO4).

The rest of the paper is organized as follows: Section 2 presents the theoretical framework, Section 3 discusses the materials and method, including the science mapping analysis framework, data collection processes and procedures, and analysis techniques. Section 4 presents and analyzes the results. The final part, Section 5, discusses the research outcomes and presents policy recommendations and suggests areas for future research.

## 2. Theoretical Background: Conversational Chatbots and Generative AI Models

AI model development, which started with the works of early pioneers like Alan Turing between 1950 and 1980, centered on rule-based and expert systems (Belgum et al., 1988; Breuel, 1990; Copeland and Proudfoot, 2000; Delipetrev et al., 2020). Those AI systems relied on predefined rules and knowledge bases to make decisions. In the 1980s, the first ideas on connectionism emphasized using neural networks to model cognitive processes (Bozinovski, 2020). In the 2000s, interest in generative models grew, which saw the use of probabilistic models, including Hidden Markov Models (HMMs) and Latent Dirichlet Allocation (LDA) models to generate data (Zeng and Zhang, 2007). The concept of generative models expanded beyond traditional statistical methods, paving the way for more sophisticated approaches, as occurred over the years. Currently, variational autoencoders (VAEs) and generative adversarial networks (GANs) proposed by Goodfellow et al. (2014) and Goodfellow et al. (2020) emerged as groundbreaking generative models focused on probabilistic generative modeling and revolutionized the field, introducing generative models that create high-quality and realistic data such as images. Vaswani et al. (2017) proposed transformer architectures that profoundly impact NLP and generative tasks. According to Kalyan et al. (2021), the generative pre-trained transformer (GPT) models and bidirectional encoder representations from transformers models possess exceptional capabilities in language understanding and generation. Recent trends focus on training models on enormous datasets and fine-tuning them for specific purposes like ChatGPT. This technique has produced cutting-edge solutions in several generating tasks. However, the growing capabilities of generative models present ethical challenges, such as biases, abuse, and the responsible deployment of AI-generated material (Chimbga, 2023).

CGAI models are designed to create new data instances resembling the training datasets, such as the generative adversarial networks (GANs), the variational autoencoders (VAEs), and transformer-based language models like GPT (Harshvardhan, 2020; Goodfellow et al., 2020; Liu et al., 2021). A transformer architecture is an instance of neural network architecture designed to capture dependencies and relationships in sequential data efficiently (Le et al., 2021; Chitty-Venkata, 2022). It plays a vital role in the success of generative models, especially in natural language processing applications. Most generative models are built using an adversarial training technique (Han et al., 2019; Lewis et al., 2020). In this case, two neural networks (one a generator and the other a discriminator) compete to enhance the quality of the created data. Several sequences of models are built to provide a conversational environment. Natural language understanding (NLU) enables such a system to interpret user input and derive meaning from it (Meera et al., 2022). These models can perform sentiment analysis, entity extraction, and intent identification. The conversational process guarantees coherent and contextually relevant interactions throughout the conversation by regulating the flow of the discussion, upholding context, and choosing acceptable responses (Meera et al., 2022).

Generative and conversational AI development involves drawing on various foundational theories from various fields, such as leveraging probability and statistical principles to model and generate new data points that resemble the training dataset (Kersting et al., 2011). Dialog systems and human-computer interaction (HCI) principles guide the design and implementation of conversational AI systems to ensure effective and user-friendly interactions. Theoretical frameworks and progress from speech processing contribute to developing conversational AI systems that can understand and generate spoken language. Similarly, the generative adversarial networks are based on game theory, where a 'generator' or and a 'discriminator' engage in a dynamic competitive interplay process to improve the generator's ability to produce realistic data (Zhang et al., 2023). The Understanding and application of these foundational theories build the essential robust and effective theoretical foundation and frameworks needed to push the boundaries of what AI can achieve in generating content and engaging in meaningful conversations.

Recent advances like ChatGPT are based on transformer architecture and represent a leap in generative and conversational AI (Ahmed, 2023). Pre-trained on vast amounts of internet text, ChatGPT demonstrates the power of large-scale pre-training for generating contextually relevant and coherent responses (Ray, 2023).

## 3. Materials and Method

### 3.1 Data Collection

The data in this study was obtained by surveying the Web of Science (WoS), a leading bibliographic database that indexes high-quality sources and scientific literature in many disciplines. As established in previous studies, WoS indexes top quality sources and publications (Jacso, 2005; de Moya-Anegón et al., 2007). The WoS platform offers users the opportunity to query the research citation database using sets of keywords built as 'query string' or search criteria (Donthu et al., 2021; Akpan, Shanker, and Offodile, 2023). The search keywords focused on the use of generative artificial intelligence (GAI), Open AI (Chatbots and ChatGPT), and human-chatbots interaction in higher education. The rationale for choosing WoS is to ensure the quality of the publications from where we extract the metadata for analysis.

The data collection occurred on Nov 30, 2023. The query returned 1922 publications, which reduced to 1,838 after filtering and final screening based on the criteria listed on Table 1. The WoS platform offers the option to obtain the data in different formats including comma-separated values (.csv), text (.txt) and other file formats. This study extracted complete records from 1,838 well screened published records as metadata in the text (.txt) for-mat, which were exported onto EXCEL for cleaning and screening. In the screening and selection process, all non-peer-reviewed documents such as meeting abstracts, notes, letters, and editorials were removed. Irrelevant records that did not address the subject of interest in artificial intelligence application in higher education and research were discarded. Table 1 presents the search terms and the query strings for data collection and the screening and selection processes and procedures (Donthu et al. 2021)

Table 1 Literature survey and data collection process: Search and retrieval, filtering, screening, and selection criteria of scientific literature on generative artificial intelligence in education.

| Activities/Focus | Criteria |
|---|---|
| **Data Source(s)** | Web of Science (WoS) Bibliographic Database. |
| | (("chatbot" OR "chatgpt") AND ("educat*" OR "teach*" OR "research" OR "e-learn*")) AND PUBYEAR = ALL YEARS, which produced articles from 2006. The search generated 2143 published documents. Nov 30, 2023. |
| | **Documents Filtering, Screening, and Selection** |
| **Filtering** | Removed 221 Documents [2143-221=1922]; Editorial Material (126); Letter (84); Meeting Abstract (3); News Item (3); Correction (2); Data Paper (2); Book Review (1) |
| **Screening** | 1922-Non-English (34) = 1,888 |
| **Final Documents Selection** | 1888-50=1838 publications that address Chatbots and ChatGPT in higher education between 2006 and 2023. Documents retrieved in text formats (.txt and .csv files) for analysis. |

### 3.2 Science Mapping and Analytics

Science mapping employs computational and analytical methods to evaluate research fields, scientists, sources and thematic structure of publications, and visualization. It involves statistical analyses of publications to track and evaluate trends and performances and map the relationships among the published documents and the contents (Donthu et al., 2021; Van Leeuwen, 2004).

The technique also employs network analysis and visualization as part of the science mapping analysis to enrich the output and results presentation (Cobo et al., 2011; Yu et al., 2017; Akpan & Akpan, 2021). The performance measures include publication trends, citation impacts, and collaboration indexes, while co-citation analysis and co-occurrence of keywords analysis to identify the research streams and themes (Donthu et al., 2021). These quantitative and descriptive bibliometric techniques help to analyze literature's conceptual, intellectual, and social structures (Van Leeuwen, 2004; Cobo et al., 2011; Kobara and Akpan, Effiom, & Akpanobong, 2023). The quantitative or evaluative methods of scientific productions help uncover the research trends, themes' dynamics, performance, and citation impact analyses.

### 3.3 Science Mapping Software

Recent advances in big data analytics and the availability of software solutions to analyze bibliographic data enhance bibliometric analysis greatly (Cobo et al. 2011; Kobara and Akpan, 2023). This study utilizes two bibliographic software packages: an R-based Bibliometrix (Aria & Cuccurullo, 2017) and VOSviewer (Van Eck & Waltman, 2010). The Bibliometrix package is embedded in the R-Studio environment (Aria & Cuccurullo, 2017). The VOSviewer appears relatively easier to produce bibliometric network analysis and visualization capabilities (Van Eck & Waltman, 2010). Further, the two bibliometric evaluation applications can handle "big data" and produce quantitative results, network analysis, and visualization (Wang et al., 2020; Akpan and Offodile, 2024).

## 4. Results and Analysis

### 4.1 Sample Description and Results Summary

The metadata analyzed in this study came from 1838 samples (N=1838). As explained in the previous section, the samples were documents published in 1025 sources and addressed topics on conversational chatbots and CGAI applications for teaching, learning, and research. The data was extracted in text format and exported onto the R-based Bibliometrix and VOSviewer software for analysis and visualization. The results obtained helped address the research objectives explained in Section 1. Table 2 presents the results summary.

Table 2 Summary of metadata from publications on Chatbots and ChatGPT in Higher Education.

| Description | Results | Description | Results |
|---|---|---|---|
| Years of Publications | 2006-2023 | *Journal Articles (1424); Book Chapters (3); Proceedings (411)* | |
| Sources | 1,025 | ***Other Documents Info:*** | |
| ***Documents Information:*** | | Document Average Age (Years)* | 1.19 |
| Total Publications | 654 | * Publications in 2023: As of Nov 30. | |
| Annual Publications growth rate % | 8.01 | ***Authors and Collaboration:*** | |
| Average citations per document | 7.51 | Authors | 6,799 |
| Total references | 59,440 | Authors of single-authored document | 167 |
| ***Documents Contents:*** | | Single-authored documents | 173 |
| Keywords Plus | 1,560 | Co-Authors per document | 4.31 |
| Author's Keywords | 4,659 | International co-authorships percentage | 26.17 |

## 4.2 Trends Analyses

### 4.2.1 Trend Analysis of Annual SLP

The results of the publication trends and citation impact on conversational chatbots and GAI (e.g., OpenAI/ChatGPT) application in education and research helps to address the first research objectives (RO1) as listed in Section 1. Figure 1 shows the annual SLP productivity trends and the earned citations covering the period (2006-2023).

The 1838 documents consisted of 1424 (77.5%) journal articles, 411 (22.3%) conference proceedings, with a negligible number (3 or 0.2%) as book chapters. The annual SLP growth rate for the during the period was 8.01% (Table 2) and a yearly average of 102. The highest publications in a single year (981 or 53.3%) occurred in 2023. The results further showed that just 4% (n=75) of the SLP occurred during 2006-2018, while the past five years (2019-2023) popularly known for increased digital transformation, fourth industrial revolution, and human-robot interaction and collaboration witnessed an exponential growth (n=1763 or 96%). Also, the last five years constitutes the trending period of CGAI such as ChatGPT generally and increased use in education and research activities.

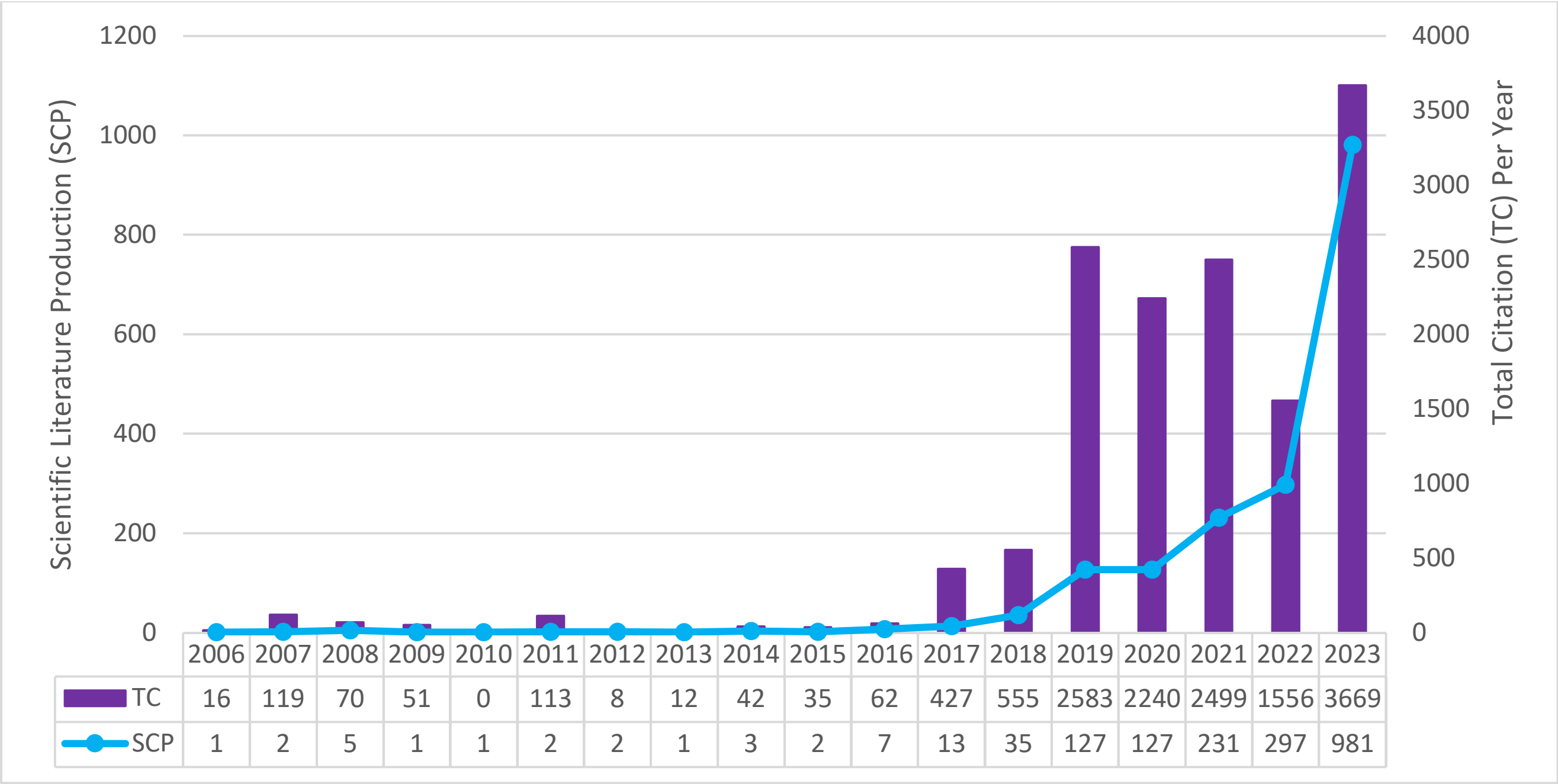

| | 2006 | 2007 | 2008 | 2009 | 2010 | 2011 | 2012 | 2013 | 2014 | 2015 | 2016 | 2017 | 2018 | 2019 | 2020 | 2021 | 2022 | 2023 |
|---|---|---|---|---|---|---|---|---|---|---|---|---|---|---|---|---|---|---|
| TC | 16 | 119 | 70 | 51 | 0 | 113 | 8 | 12 | 42 | 35 | 62 | 427 | 555 | 2583 | 2240 | 2499 | 1556 | 3669 |
| SCP | 1 | 2 | 5 | 1 | 1 | 2 | 2 | 1 | 3 | 2 | 7 | 13 | 35 | 127 | 127 | 231 | 297 | 981 |

Figure 1. Annual scientific literature publications (SLP) and citation impact trends on generative artificial intelligence and open artificial intelligence (Chatbots and ChatGPT) in higher education (2006 – 2023).

#### 4.2.2 Multidisciplinary Perspective of CGAI in Education and Research

The literature highlights the multidisciplinary application of CGAI (Chatbots/ChatGPT) impacts and challenges in education and research. This Section examines different fields or disciplines that use CGAI the most for teaching, learning, and research. We adopt the WoS subject/research categories, identifying over 50 multidisciplinary areas (Leydesdorff et al., 2013; Martín-Martín et al., 2018; Akpan and Akpan, 2021). However, we merged the closely related fields in this study and reduced the classification to 21 different areas (Figure 1). For example, we combined some computer science areas (software, hardware, systems, and methods) to areas to form one classification. The results identify the subject areas and application domains addressing CGAI in education (Figure 2).

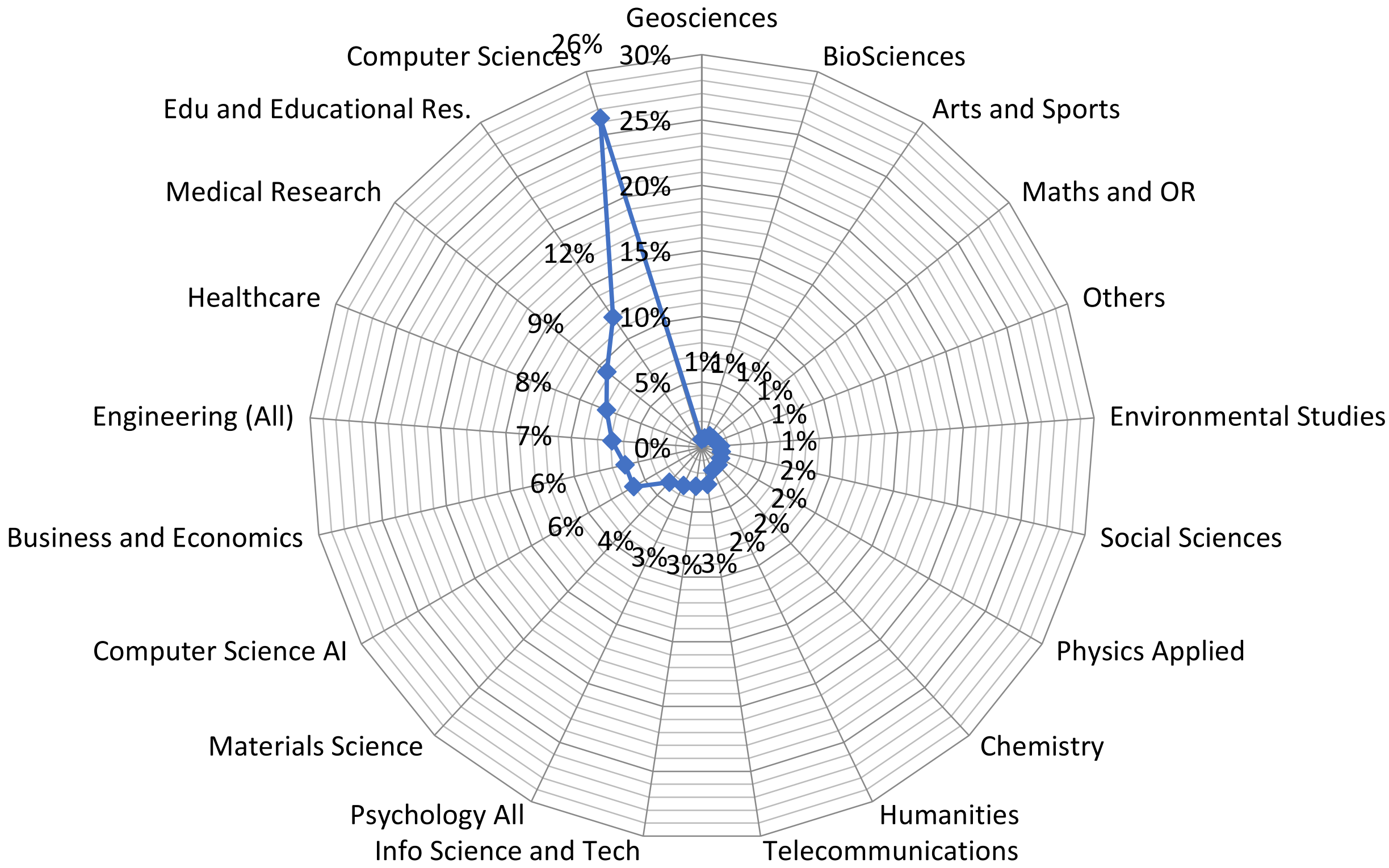

Figure 2: Multidisciplinary fields and application domains of Chatbot/ChatGPT

The most popular academic disciplines and research fields adopting ChatGPT and other CGAI for teaching, learning, and research include "computer science multidisciplinary" (26%), "education and educational research" (12%), and "medical research" (9%). Others are "healthcare" (8%) and "engineering multidisciplinary" (7%), while "business and economics" and "computer science artificial intelligence" both make up 6%. About six (6) academic disciplines recorded minimal CGAI use, including "environmental studies," "mathematics and operations research," and "arts and sports," among others. The listed field with minimal application contributed just 1% of the SLP.

**4.2.3 Analysis of SLP Usage**

The frequent usage of GAI (Chatbot and ChatGPT) in education publications indicates an active research genre in the past six months (June to November 2023) or 180 days (U1) and the past nine years, 2015-2023 (U2). The results show that the published documents in the past nine years have been accessed 46,088 and 80,185 for U1 and U2, respectively (Figure 2). However, the results do not show the purpose of use, although the potential users can include educators, students, researchers, industry practitioners, and others.

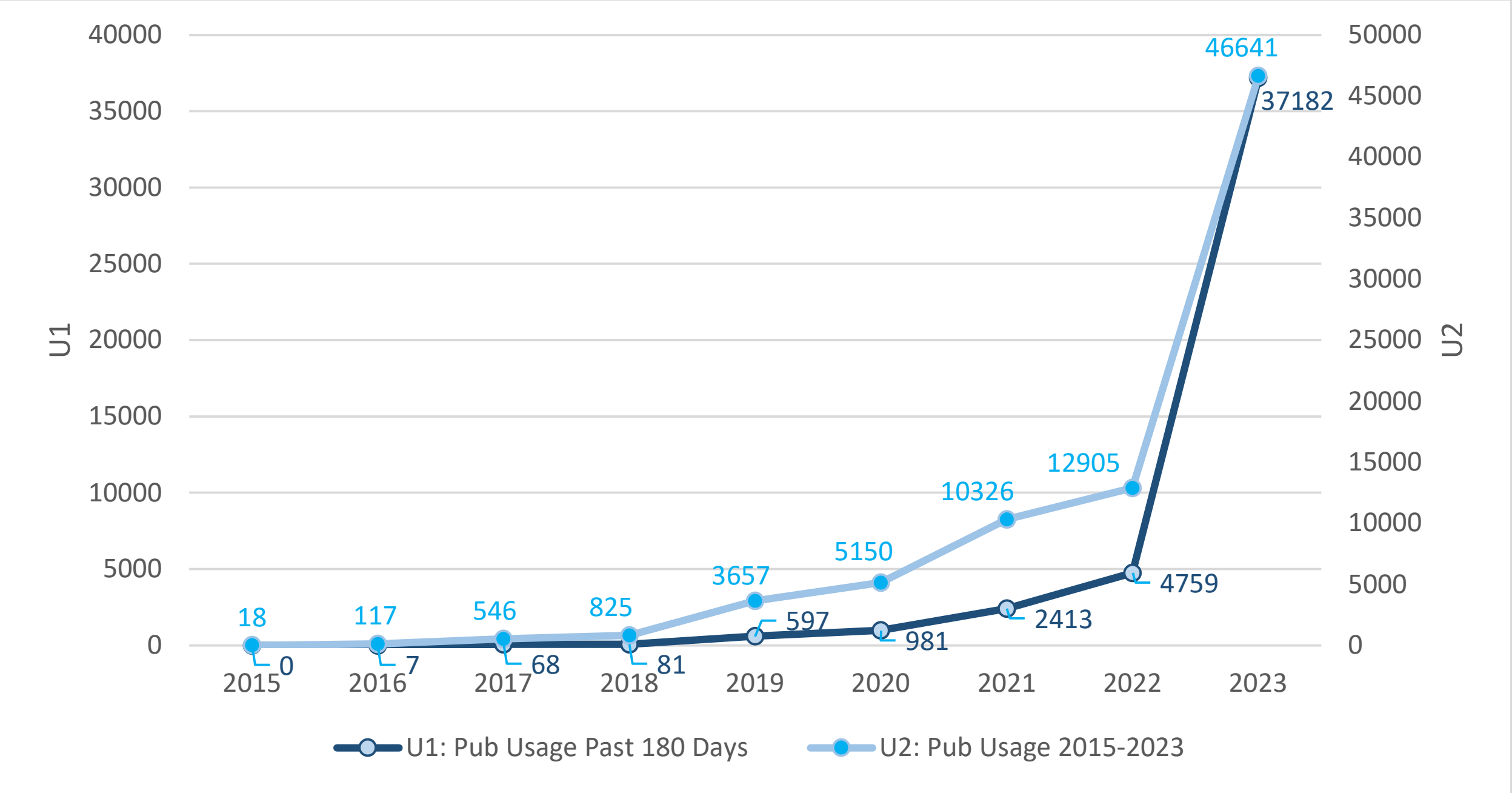

Figure 3. Usage frequency of scientific literature on generative artificial intelligence (Chatbot and ChatGPT) in Education in the last six months (U1) and from 2015 to 2023 (U2) based on WoS bibliographic data.

The current year (2023), which also recorded the most publications (over 53%) and recorded the most SLP usage (U1=37,182 times; U2=46,641 times, respectively). The trend analysis for U1 and U2 also shows a consistent increase in SLP use across all years, with recent publications used more often, meaning that recently published articles received higher usage counts than the older ones. For example, the 981 articles published in 2023 have been used 37,182 times in the last six months and 46,641 times during the year (2023), from 4,759 and 12,905 SLP usage in 2022, representing an increase of over 681% (U1) and 261% (U2), respectively. Furthermore, the usage trends for U1 and U2 mirror the increase in the literature publications in recent years (2015 to 2023), indicating an active research discipline (Figures 1 and 2). Both U1 and U2 followed a similar upward trend (Figure 2). However, the current year (2023) shows a steep upward slope.

### 4.3 Thematic Structure, Use Cases, Benefits and Limitations of CGAI in Education and Research

#### 4.3.1 Mapping the Scientific Landscape of CGAI Application

A science mapping of the scientific literature landscape on CGAI application in educational services operations and research, including teaching, learning, and creative activities reveals the thematic structure, use cases, and SWOT analysis of chatbot-human interaction. The results also highlight the use cases, and the benefits and challenges, which help to address the second research objective (RO2).

The text analytics using the R-Bibliometrix identifies 4661 unstemmed author keywords with a total word frequency (f) of 10057 (the number of times each keyword re-occurs in the dataset). The sample is stratified into two categories to enhance an exhaustive analysis as follows:

- Prominent themes, defined by prominent keywords with frequency (f) of ten or more, (f >= 10).
- Emerging themes, having low frequent keywords of less than ten, (f<10).

#### 4.3.2 Prominent Research Themes

The eminent research themes on CGAI are determined by the frequency of occurrences of the terms during the period covered in this study. We define the terms with ten (10) or more frequencies (f >= 10) as prominent. The results using the Bibliometrix application in the R-studio environment show seventy (70) unique but unstemmed keywords (as a pointer to the research themes in this study) with a total frequency of 3564, constituting 34.4% of the total. After keyword stemming, the theme frequency, however, reduced to twenty-two (22). Word stemming combines terms with the same meaning even though spelled differently (e.g., "AI," "artificial intelligence," "artificial intelligence-AI"). The concepts of word stemming as a text analytics method are available elsewhere (Xu and Croft, 1998; Bullinaria and Levy, 2012; Akpan et al., 2023). The text analytic software application considers terms with different spellings as unique despite being similar in meaning. Figure 4 presents the prominent themes after stemming.

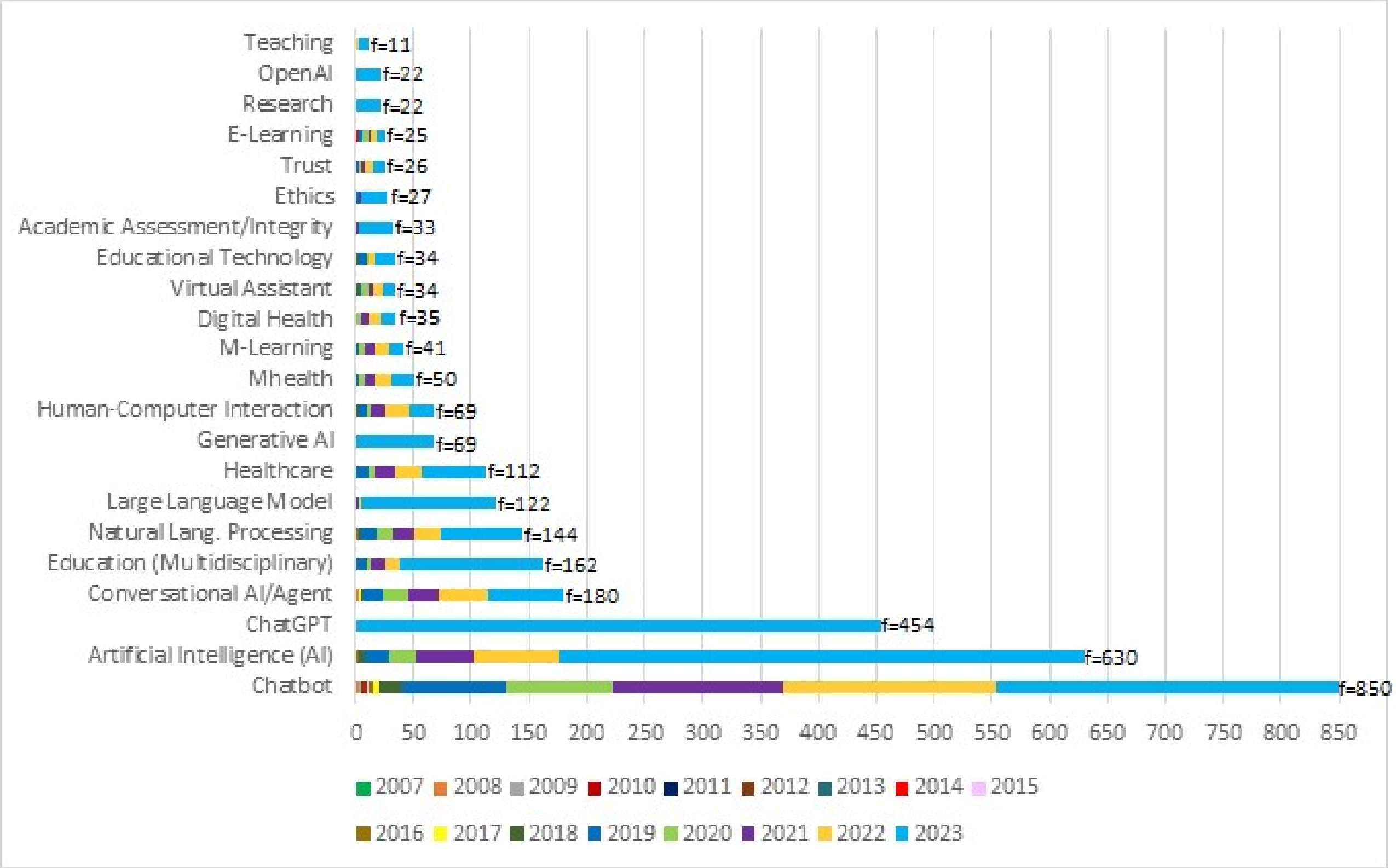

Figure 4. Yearly trends of eminent research themes on CGAI application in education and creative activities.

The first part includes the AI concepts and CGAI agents, such as "Chatbot" (850 or 8.5%), "artificial intelligence" (630 or 6.3%), "ChatGPT" (454 or 4.5%), "conversational agent" (180 or 1.8%) and "virtual assistant" (34 or 0.34%) of total word frequency. The main areas of CGAI application includes “education (multidisciplinary” (162 or 1.62%), “healthcare" (112 or 1.1%), “m-health/digital health” (“85” 0.85%) and "human-computer interaction" (69 or 0.69%). The number in the parenthesis represents the frequency as explained earlier.

### 4.3.3 Emerging Themes

In this study, the research themes that occurred 1-9 times (f<10) is classified as emerging themes. The results of the text analytics identified 4591 unstemmed emerging themes with co-occurrence frequency of 6493 times (i.e., f=6493). This constitutes about 64.6% of the total word frequencies in the metadata based on the keywords as the unit of analysis. This data is utilized to create a network map and visualize the thematic structure of the scientific literature landscape on the application of CGAI in education and scholarly activities using VOSviewer application (Waltman & van Eck, 2013). The process also includes keywords stemming (explained in Section 4.3.2). Nugatory terms or words that do not convey contextual meanings, such as author affiliations are also removed (Waltman & van Eck, 2013; Akpan and Offodile, 2024).

The co-occurrence of the author keywords and the interconnectedness are represented by means of nodes (circles), edges (lines), and node labels to produce the network map and visualization of the thematic structure of CGAI literature landscape on a network map. The nodes' sizes and labels depend on the number of occurrences of specific keywords, which are pointers to the research topics or subjects. The bigger the node, the more frequently the word occurred (Cobo et al., 2011; Kobara and Akpan, 2023). The results show color-coded themes categorized into clusters identifiable by color-coded nodes, implying that nodes with the same colors belong to the same cluster, ordered by year of publication (Figure 5).

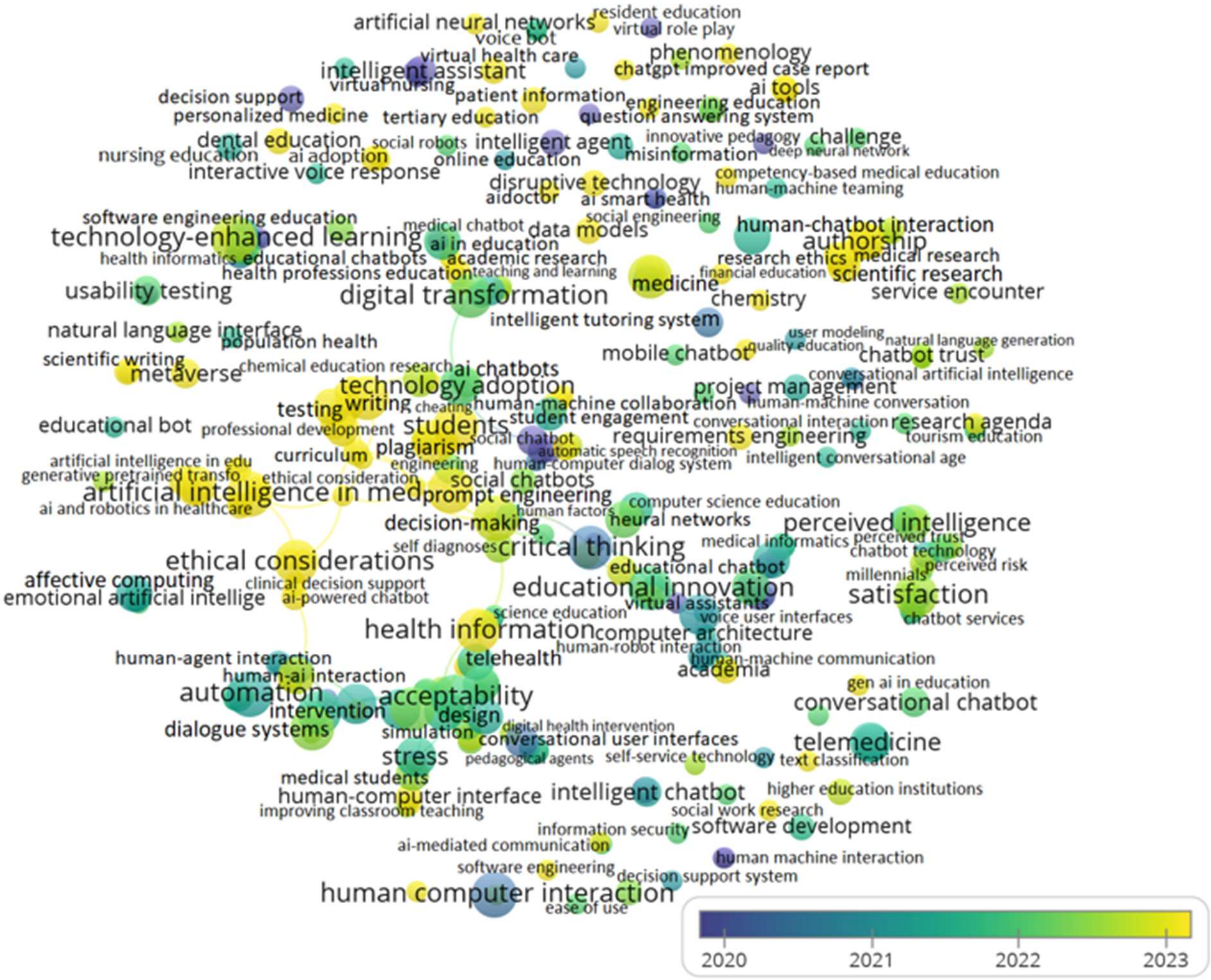

Figure 5. Visualization of the thematic structure of CGAI application in education and research activities

We classify the clusters into two categories based on years of publication: the newest clusters (2020-2023) and the older years (2006-2019), indicated by purple color-coded themes (Figure 5). The yearly trends classification corresponds to the analysis of the prominent themes in the previous Section 4.3.2. Some of the keywords captured in each Cluster in the emerging themes category depicted as words/phrases include the following: 1. **Bright-Yellow Cluster - 2023** ("disruptive technology," "ChatGPT improved case report," "artificial intelligence in med," "health education," "gen ai in education," "academia," "financial education," "students," "testing," "writing," "curriculum," "educational chatbot," "ethical consideration," "scientific research," "authorship," "plagiarism," "research ethics," "ai-mediated communication," "automatic speech recognition," "web-based learning," among others); 2. **Light-green Cluster –**

**2022** ("digital transformation," "technology adoption," "acceptability," "educational innovation," "AI in education," "conversational chatbot," "medical chatbot," "telemedicine," "health literacy," "engineering education," "software development," "chatbot services," "mobile chatbot," "voice assistant," and more); 3. **Green Cluster – 2021** ("human-chatbot interaction," "human-machine teaming," "student engagement," "customer experience," "information security," conversational artificial intelligence," "intelligent chatbot," "usability testing," and more); 4. **Purple Cluster – 2020** ("knowledge base," "AI smart health," "dialog system," "question answering," "attention mechanism," conversational systems," "intelligent chatbot," and more); 5. **Dark-Purple Cluster – 2006-2019** (human-machine interaction, "intelligent assistant," "disaster management," "decision support," "intelligent personal assistant," "natural language understanding," "conversational interfaces," "peer assessment").

The themes and keywords point to the thematic structure of CGAI application in teaching, learning, life application, and research/scholarly activities. The following sections examine further analysis of the prominent and emerging themes, highlighting the CGAI use cases, benefits, and limitations in education and research.

#### 4.3.4 Multidisciplinary Perspective of CGAI Use Cases in Education and Research

Mapping the research landscape on CGAI in this study identifies the CGAI contexts, use cases, strengths, and limitations and identifies the gaps in educational policies needed to address the ongoing disruptions in educational services and research activities (Amankwah-Amoah et al., 2024; Farina and Lavazza, 2023; Ahmad et al., 2023). The results show that Chatbots (e.g., ChatGPT and other bots) have significantly disrupted education in many disciplines (Mondal et al., 2023; Dwivedi et al., 2023). Table 3 summarizes the ongoing CGAI applications and highlights the benefits and impacts in multidisciplinary fields.

Table 3 Multidisciplinary CGAI use cases, strengths, and limitations

| AI Agent | Use Cases | Benefits | Limitations | References |
| --- | --- | --- | --- | --- |
| ChatGPT | AI agent-human collaboration in learning. | Digital writing and composition; personalized learning. | Academic integrity concerns. | Perkins, 2023; AlAfnan et al., 2024; Kooli, 2023. |
| ChatGPT | Multidisciplinary: Students' use for tests/exams; Scientific research. | Aiding in exams, generates responses and can write essays for college students. | Cheating, issues relating to attribution and plagiarism. | Farina and Lavazza, 2023; Ahmad et al., 2023; Heng et al., 2023; Alkaissi and McFarlane, 2023 |
| ChatGPT | Learning about effects of multiple drugs interaction. | Predicting multiple medication interaction. | Lack of human interaction. | Juhi et al., 2023; Ahmad et al., 2023 |
| ChatGPT | Teachers use to evaluate curricula, syllabi, and formal assessments. | Generating course material, suggests content; knowledge and skills to learn | | Condrey, 2023; Martinez et al., 2021; Chiu et al., 2023 |
| ChatGPT | Education: Evaluated Academic Integrity Policy against CGAI | The evolving education landscape given the AI technology revolution needs policy reviews. | Most policies lack explicitness of technology mention. | Perkins and Roe, 2023 |
| ChatGPT | Computer Programming & Code Generation | The AI Chatbot can generate code and helps to complete programming tasks. | | Piccolo et al., 2023; Merow et al., 2023; |
| Online Educational Chatbot | Implements dialogue system with ability to pace and guide materials to inspire learners. | Offers context recognition, emotion management; improves learners' emotional confidence and overall learning experience. | To strengthen attention mechanisms; emotional connections with learners. | Zhang et al., 2022; Fadhil and Villafiorita, 2017 |
| Interactive Learning Chatbot | Chatbot for interactive learning during COVID-19. | Provided digital healthy-coping intervention and support to distressed university students. | | Gabrielli et al., 2021; Balderas et al., 2023 |
| AI Medical Chatbot for Learning; ChatGPT (GPT4) | Family medicine residency progress test; Medical education. | Potent medical education tool; able to create exam questions; a resource for medical knowledge. | | Huang et al., 2023; Oh et al., 2023; Heng et al., 2023 |
| -DetectGPT; -GPTZero; -AICheatCheck | Detects scientific writing by human, Chatbot, or a combination of both. | Able to detect the authorship of scientific text; ensure integrity of scientific writing. | | Abdalla et al., 2023; Adeshola and Adepoju, 2023; Ghani (2023) |
| Web-enabled language learning Chatbot | E-learning: English as foreign language | Posses good virtual learning capability | Students' loss of interest in Eng. lang. learning. | Fryer et al., 2017; Huang et al., 2022 |
| Chatbot & Humanoid Robot | Supporting science education (chemistry) | Enhances students' concentration. | | Matsuura and Ishimura, 2017 |
| Conversational Chatbot | Online learning; | Enhances students' interaction with online courses. | | Song, Oh, and Rice, 2017; Rasheed et al., 2019 |
| Conversational Chatbot & Human-Chatbot Interaction | A digital personal assistant; offers technical support/ customer service | Engage in conversations with human users; high perceived humanness and engagement. | Continuous improvement of the humanness. | Schuetzler et al., 2020; Pizzi et al., 2021; Pillai and Sivathanu, 2020. |
| Conversational Chatbot | Education: learning and administrative purposes | Dialog system supporting student learning and for administering exam | Limited mobile learning capability. | Rasheed et al., 2019 |
| Mobile Learning Chatbot | Educational chatbot for persons with disabilities | Can be utilized in multi-purpose device; offers voice or text interaction. | | Mateos-Sanchez et al., 2022 |
| Virtual Assistant – Chatbot for Learning | Science Education (Physics) | Increased the conceptual understanding of Newton's laws. | Should improve interaction features. | Aguilar-Mejía and Tejeda, (2020). |
| Question answering chatbot | Offers answers to users' questions on biomedical topics. | Biomedical research: offers answers to users' questions on biomedical concepts. | Domain specific chatbot, which can be limiting. | Xygi et al., 2023 |
| ChatPal, Chatbot for Mental-health (MH) analytics | ChatPal promotes positive mental well-being among rural dwellers | The MH Chatbot usage logs and patterns offers insight into users' mental well being. | | Booth et al., 2023; Potts et al., 2023 |
| Human-Chatbot Interaction | Multilingual Conversational Chatbot | Conversational Chatbot suitable for multicultural community with many languages and accents. | | Adeniyi et al., 2022; Chen et al., 2023 |
| Human-Computer Interaction | Health Chatbot acceptability among general practitioners. | Chatbots least acceptable as a consultation source for severe health issues. | Privacy issues, such as personal information. | Miles et al., 2021 |

Figure 2 (Section 4.2.2) identified more than twenty (20) application domains of CGAI based on Web of Science classification. The CGAI use cases in educational multidisciplinary services, including computer science (computer programming and human-computer interaction), medical/healthcare, biomedical engineering, and business. The section also discusses the strengths and limitations of CGAI in the mentioned domains. The analyses help to address the second research objective (RO2).

The use of CGAI in teaching and academic instruction cuts across several disciplines. The use cases include generating course material for teachers and suggesting what knowledge, skills, and other contents (Dwivedi et al., 2023). For example, Yilmaz et al. (2023) offered sample use cases of using ChatGPT to generate a course syllabus for an undergraduate statistics course and other valuable suggestions. ChatGPT also guides instructors in generating exercises, quizzes, and scenarios for student assessment (Geerling et al., 2023; AlAfnan, 2023). Using chatbots and advanced conversational AI models like ChatGPT in academic instruction and student learning has several implementations, widespread adoption, and opinions depending on the user.

Similarly, students also utilize Chatbots and ChatGPT for learning and academic writing, such as personalized learning support and answering questions related to course materials, assignments, or general queries. Other benefits include offering immediate answers and feedback on assignments and improving work (Wang and Chen, 2023; Rawas, 2023; Opara et al., 2023). Most educational Chatbots, such as ChatGPT, facilitate virtual Q&A sessions, allowing students to ask questions about anything at any time and receive instant responses (Gabrielli et al., 2021; Mane et al., 2023).

Educational services Chatbots are also effective in language learning or as translators, facilitating engagement in conversations, helping students practice and improve language skills, and offering vocabulary suggestions, grammar corrections, and language-specific guidance. In international or multilingual academic settings, ChatGPT helps translate language and facilitates communication among students and instructors (Jiao et al., 2023; Yilmaz et al., 2023; Stap et al., 2023). In addition, ChatGPT can help students generate written content, offering prewritten essays and research papers and substituting the overall writing process (Devi et al., 2023; Morrison et al., 2023; Norris et al., 2023). Ahmed et al. (2023) reported that ChatGPT has also been used as a virtual tutor to explain complex concepts, helping students grasp difficult subjects and topics (Lo, 2023; Sok and Heng, 2023). Some students have used ChatGPT to assist in preparing for interviews by simulating interview scenarios and questions, offering practice for articulating thoughts and responding to questions (Mellor, 2023; Iskender, 2023).

In addition to teaching and learning, CGAI in education is also used for academic writing and scientific research, which have demonstrated high capability and versatility in generating human-like text and assisting researchers. According to Lund et al. (2023), ChatGPT generates assessments, provides virtual personal learning, and facilitates research. Further, ChatGPT can create scientific papers from title to introduction, backgrounds, articles, and creative writing, aiding researchers in exploring language generation capabilities. Lund et al. (2023) argue that GPT can write hundreds-word essay at the caliber of a professional researcher in less than a minute. It is more efficient when the topic is divided into smaller parts than when GPT is used to write each subsection of the post. Some researchers used it to generate synthetic data for various purposes. Researchers can leverage ChatGPT to research information, while others use it to develop and enhance question-answering systems. According to Hosseini et al. (2023) and Stokel-Walker and Van Noorden (2023), ChatGPT can be used to check and fix typographical and grammatical errors in research works.

Some researchers use ChatGPT as a collaborator in research, depending on the content generated in the paper. Many researchers report using ChatGPT as a text summarization or core content abstracting tool from articles (Hosseini et al., 2023). The prompt given to ChatGPT can be modified to generate concise and informative summaries of longer texts and abstract information. Haman and Skolnik (2023) and Zamfiroiu et al. (2023) tried using ChatGPT to conduct a literature review, showing that the GPT bot has real potential in academic writing and publishing. However, educational institutions must develop policies to address the threats against academic integrity concerns and plagiarism (Farina and Lavazza, 2023; Ahmad et al., 2023; Alkaissi and McFarlane, 2023).

### 4.3.5 Conversational Chatbot and Generative AI Applications in Operational (Operations) Research

The OR discipline and allied industry operations are some of the prominent areas adopting artificial intelligence to improve production and operations processes and performance efficiency (Fethi & Pasiouras, 2010). Generally, OR offers diverse methods and techniques to optimize operational activities in different industries and economic sectors through developing mathematical, computational, and analytical models of complex systems that provide insights and solutions to enable stakeholders to make informed decisions and improve efficiency and productivity (Arnaout & Maatouk, 2010; Fethi & Pasiouras, 2010). As OR employs computational algorithms and data analytics methodologies to solve complex business and operations problems in diverse economic sectors (Doumpos et al., 2023; Kanbach et al., 2024), integrating GenAI into decision support systems can ensure better outcomes. Table 4 itemizes the use cases of CGAI in OR to improve business operations and process innovations.

Table 4 Potential uses or benefits and shortcomings of CGAI in operational/operations research

| Use Cases | Description of Potential AI Integration in OR | Limitations/Further Research Areas |
|---|---|---|
| Model Creation | OR methods and techniques generally involve the creation of mathematical or computer models of complex systems for decision-making. The OR analyst can feed a description onto CGAI (ChatGPT) to create the model (Kanbach et al., 2024). | Reliability issues: Does the model created by CGAI accurately represent the intended system? |
| Decision Support Systems | Another OR objective is to aid decision-making. Decision support system is any computer program that helps to achieve this. Using data analytics capability, CGAI can analyze scenarios based on synthetic data as proxy to real world data and produce solutions to aid informed decision-making (Benaim et al., 2020; Akpan and Brooks, 2014; Bietsch et al., 2023; Jacobsen, 2023). | CGAI can be trained to generate large synthetic data. However, such data do not always mimic real world scenarios. |
| Optimization | Operations or process optimization is at the center of OR objective (performance efficiency, costs minimization or profit maximization). ChatGPT can produce algorithms (e.g., generative adversarial networks - GANs) can produce optimal solutions for complex operational systems (Gui et al., 2021; Amorim-Lopes et al., 2021). | Further studies can be conducted to examine the challenges of GANs (e.g., Saxena et al., 2021). |
| Simulation | CGAI can create models that simulates complex systems to produce optimized solutions for different scenarios either to improve operations, performance efficiency, and process innovations (Doumpos et al., 2023; Akpan et al., 2024). | How valid and credible are simulation models created by CGAI? |

However, Since OR and allied fields are not immune to the identified challenges associated with CGAI/GenAI adoption, there is a need to seek ways to implement systems to check for any forms of abuse or inadequate implementation toward realizing the full potential and benefits identified in Table 4.

### 4.4 Limitations and Concerns of CGAI

The previous Section discussed several benefits of CGAI (ChatGPT and other AI chatbots) in academia but also raised several concerns. There is increased attention to ethical considerations, bias mitigation, and academic integrity concerns. Issues relating to misuse, privacy, and authenticity remain notable challenges and concerns. The CGAI outputs often reflect biases in the training data regarding content generation. Rozado (2023) and Motoki et al. (2023) reported using several instruments to test ChatGPT biases in answering political questions. The consistent response is that ChatGPT prefers left-leaning viewpoints, although more evidence is needed, and further clarification is needed to buttress this point. Chen et al. (2023) discovered that ChatGPT has an apparent mathematical or probabilistic basis; it demonstrates a wide range of human biases, particularly in complex, ambiguous, and implicit problems. Both probability weighting and conjunction bias are present, and the choice of reference, the salience of expected regret, and the framing affect its preference. This leads to biased responses that may unintentionally perpetuate or reinforce existing stereotypes and discrimination. Sometimes, ChatGPT may generate responses that are factually incorrect, unprecise, or inaccurate (Kocon et al., 2023). In academic settings, where accuracy is crucial, reliance on some erroneous information can be misleading and contribute to spreading misinformation.

From an ethical perspective, there is potential misuse of ChatGPT for academic dishonesty, such as generating content for assignments or exams, raising ethical concerns (Sullivan et al., 2023; Perkins et al., 2023; Bin-Nashwan et al., 2023). Reviewing the potential of compromising academic integrity, educational institutions must establish guidelines to prevent inappropriate use. This also raises the question of the quality of CGAI training students receive without thorough learning by generating information from AI platforms with no creativity and no intellectual progress from the students (Padillah, 2023). CGAI-directed learning can impact students' quality of life, potentially affecting the future of society. Adeshola & Adepoju (2023) suggested that educational establishments can lessen the disruptive effects of this technology and uphold academic integrity by creating tests with minimal AI-generated content and through explicit regulations and guidelines.

Another concern raised when using ChatGPT is the inadequate handling of sensitive topics. ChatGPT may not appropriately handle sensitive or controversial topics, potentially leading to inappropriate or offensive responses (Sebastian et al., 2023, Kalla and Kuraku, 2023). The trend over the years is people's overreliance on technology (Baxter et al., 2017; Nyholm et al., 2023; Hayes et al., 2022). Depending heavily on ChatGPT's capacity to generate text can result in an overreliance on technology at the expense of human interaction, creativity, and novelty. Students may miss out on the interpersonal aspects of learning, such as mentorship, collaborative discussions, and direct interaction with instructors (Tlili et al., 2023). The missing social and emotional aspects of education can impact student engagement and motivation, hence learning quality. The use of ChatGPT involves sharing user inputs and raising concerns about data privacy and user security. Wu et al. (2023) argue that people can unintentionally become victims of CGAI misuse. Thus, everyone must learn how to protect themselves from these Chatbots. Gupta et al. (2023) and Kamalov et al. (2023) recommend stringent measures to protect sensitive student information and enforce compliance with data protection regulations.

## 4.5 Intellectual Structure of CGAI Scientific Literature

The third research objective (RO3) evaluates the intellectual structure of the scientific literature landscape of CGAI as addressed in this Section.

### 4.5.1 Citation Impact Trend Analysis

The results summary (Table 2) shows an average of 7.51 citations per document and 14,057 over the 18 years covered in this study. Further analysis showed that about 59.5% of the articles earned at least one or more citations, indicating a solid impact, reviewing that 96% of the publications occurred in the past five years. Also, about 41% of the publications earned between 1-9 citations; 18.1% received between ten and fifty (10<=TC>=50), while less than 1% earned 100 or more citations. This result is significant given that over 53.3% of the publications occurred in the current year and over 96% in the past five years. Further, over 40% of the SLP did not earn any citations, which can also be attributed to the fewer citable years (CY). Thus, 96% of the publications has four or less citable years (CY<5). Table 3 presents the complete citation structure of the publications (N=1838).

The reported citation counts are based on the WoS bibliographic data. The citation count on Google Scholar (scholar.google.com) is often higher because the former limits the count to publications in sources that meet its quality criteria and indexed in its literature database (Jacso, 2005; Akpan and Offodile, 2024). For example, Smutny and Schreiberova (2020), which investigated chatbots for learning, earned 152 citations as of Nov 30, 2023, based on WoS data but over 499 citations on Google Scholar.

Table 3. The spread of earned citation impact of SLP on CGAI application in education and research

| Year | >=200 | >=100 | >=50 | >=10 | >=1 | Sub-Total | NC | SLP | TC | % Cited |
|---|---|---|---|---|---|---|---|---|---|---|
| **2006** | 0 | 0 | 0 | 1 | 0 | 1 | 0 | 1 | 16 | 100% |
| **2007** | 0 | 0 | 0 | 0 | 1 | 1 | 0 | 2 | 119 | 50% |
| **2008** | 0 | 0 | 1 | 0 | 2 | 3 | 2 | 5 | 70 | 60% |
| **2009** | 0 | 0 | 1 | 0 | 0 | 1 | 0 | 1 | 51 | 100% |
| **2010** | 0 | 0 | 0 | 0 | 0 | 0 | 0 | 1 | 0 | 0% |
| **2011** | 0 | 0 | 1 | 1 | 0 | 2 | 0 | 2 | 113 | 100% |
| **2012** | 0 | 0 | 0 | 0 | 2 | 2 | 2 | 2 | 8 | 100% |
| **2013** | 0 | 0 | 0 | 1 | 0 | 1 | 0 | 1 | 12 | 100% |
| **2014** | 0 | 0 | 0 | 1 | 1 | 2 | 1 | 3 | 42 | 67% |
| **2015** | 0 | 0 | 0 | 2 | 0 | 2 | 0 | 2 | 35 | 100% |
| **2016** | 0 | 0 | 0 | 3 | 2 | 5 | 2 | 7 | 62 | 71% |
| **2017** | 0 | 3 | 1 | 4 | 6 | 14 | 0 | 13 | 427 | 108% |
| **2018** | 1 | 0 | 0 | 13 | 16 | 30 | 5 | 35 | 555 | 86% |
| **2019** | 1 | 5 | 12 | 33 | 56 | 107 | 20 | 127 | 2583 | 84% |
| **2020** | 0 | 3 | 9 | 47 | 47 | 106 | 21 | 127 | 2240 | 83% |
| **2021** | 0 | 0 | 14 | 57 | 101 | 172 | 59 | 231 | 2499 | 74% |
| **2022** | 0 | 0 | 2 | 45 | 159 | 206 | 91 | 297 | 1556 | 69% |
| **2023** | 0 | 4 | 5 | 78 | 352 | 439 | 542 | 981 | 3669 | 45% |
| Totals | **2** | **15** | **46** | **286** | **745** | **1094** | **745** | **1838** | **14057** | |
| % | **0.11** | **0.81** | **2.5** | **15.6** | **41** | **59.5** | **40.5** | | | |

*NC, TC, and NP denote no citation (publications that did not earn any citations as of Nov 30, 2023), total citations, and total publications per year, respectively, based on WoS data.*

### 4.5.1 Most Cited Documents

Table 4 presents twenty most cited articles, themes, and research focus, and the earned citations (total, average, and normalized citation values) based on the WoS bibliographic data, and the publishing sources. The three most cited themes include "challenges and opportunities with social chatbots," "chatbots and conversational agents in mental health," and "ChatGPT utility in healthcare education, research, and practice." Others include "social cues for conversational agents," "chatbot for learning," and "simulating children's role in COVID-19 transmission." Table 4 presents the complete list of titles and themes and the citation. However, as expected, most highly cited topics were published in 2020 with at least three citable years. Also, documents with less than one citable year (CY<1) have more uncited topics (Tables 3 and 4).

Table 4. Twenty most cited articles and themes on generative and open AI (Chatbot and ChatGPT) in education and research.

| S/No | Reference | Focus | Journal | TC | AVC/ Year | NTC |
|---|---|---|---|---|---|---|
| 1 | Shum, He, & Li, (2018) | Challenges and opportunities with social chatbots. | Frontiers of Information Technology & Electronic Engineering | 241 | 40.17 | 15.2 |
| 2 | Vaidyam et al., 2019. | Chatbots and conversational agents in mental health. | The Canadian Journal of Psychiatry | 234 | 46.8 | 11.51 |
| 3 | Sallam, (2023). | ChatGPT utility in healthcare education, research, and practice. | Healthcare | 178 | 178 | 51 |
| 4 | Feine et al., 2019. | Social cues for conversational agents. | International Journal of Human-Computer Studies | 159 | 31.8 | 7.82 |
| 5 | Smutny & Schreiberova, 2020. | Chatbot for learning. | Computers & Education | 152 | 38 | 8.62 |
| 6 | Lee, Bubeck, & Petro, 2023 | Benefits, limitations, and risks of chatbot in medicine. | New England Journal of Medicine | 152 | 152 | 43.55 |
| 7 | Fryer, 2017 | Comparing chatbot and human in language learning course. | Computers in Human Behavior | 113 | 16.14 | 3.44 |
| 8 | Kerlyl, Hall, & Bull, 2006 | Bringing chatbots into education. | Intl. conf. on innovative techniques & app of AI | 110 | 6.47 | 1.85 |
| 9 | Alkaissi, & McFarlane, 2023 | Artificial hallucinations in ChatGPT and scientific writing. | CUREUS | 103 | 103 | 29.51 |
| 10 | Hoermann et al., 2017 | Synchronous text-based dialogue systems in mental health interventions. | Journal of medical Internet research | 102 | 14.57 | 3.11 |
| 11 | Rapp, Curti, & Boldi, 2021 | The human side of human-chatbot interaction. | International Journal of Human-Computer Studies | 93 | 31 | 8.59 |
| 12 | Fryer, Nakao, & Thompson, 2019 | Chatbot learning partners: Connecting learning experiences. | Computers in human Behavior | 92 | 18.4 | 4.52 |
| 13 | Schuetzler, Grimes, & Scott Giboney, 2020 | Chatbot conversational skill on engagement and perceived humanness. | Journal of Management Information Systems | 77 | 19.25 | 4.37 |
| 14 | Skjuve et al., 2021 | My chatbot companion-a study of human-chatbot relationships. | International Journal of Human-Computer Studies | 76 | 25.33 | 7.02 |
| 15 | Cotton, Cotton, & Shipway, 2023 | Chatting and cheating: Academic integrity in the era of ChatGPT. | Innovations in Education and Teaching International | 70 | 70 | 20.06 |
| 16 | Shumanov, & Johnson, 2021 | Making conversations with chatbots more personalized. | Computers in Human Behavior | 64 | 21.33 | 5.91 |
| 17 | Tlili et al., 2023 | ChatGPT as a case study of using chatbots in education. | Smart Learning Environments | 61 | 61 | 17.48 |
| 18 | Pérez, Daradoumis, & Puig, 2020 | Rediscovering the use of chatbots in education. | Computer Applications in Engineering Education | 60 | 15 | 3.4 |
| 19 | Huh, 2023 | Comparing ChatGPT's knowledge and capability to med students? | J Educ Eval Health Prof. | 60 | 60 | 17.19 |
| 20 | Huang, Hew, & Fryer, 2022 | Chatbots for language learning: What is the usefulness? | Journal of Computer Assisted Learning | 60 | 30 | 11.46 |

#### 4.5.2 Eminent Sources

The sampled documents in this study were published in 1025 sources, giving an average of 1.6% publications per journal/proceedings. Table 5 shows the twenty most eminent sources regarding the citation impact and number of published documents. The twenty sources published 314 (17.1%) of the documents, earning 3980 (28.3%) citations out of 14,057. The high impact and eminent journals publishing conversational and GAI applications in education cut across many disciplines in computer science, healthcare, technology, engineering, and business. However, this is not surprising reviewing the multidisciplinary perspective of this study. The top three based on citation impact and scientific literature productivity include "Journal of Medical Internet Research," "Computers in Human Behavior," and " CUREUS Journal of Medical Science." Others are " Sustainability," "Applied Sciences," and " Education and Information Technologies." Table 5 presents the complete list of eminent journals and conference proceedings.

Table 5. The top 20 sources with most publications on Chatbot and ChatGPT, and citation impacts.

| Element | TC | NP | AVTC | Pub. Yr. |
|---|---|---|---|---|
| Journal of Medical Internet Research | 989 | 44 | 22.5 | 2017 |
| Computers In Human Behavior | 459 | 16 | 28.7 | 2017 |
| CUREUS Journal of Medical Science | 278 | 47 | 5.9 | 2020 |
| Sustainability | 146 | 18 | 8.1 | 2020 |
| Applied Sciences-Basel | 165 | 30 | 5.5 | 2019 |
| Education and Information Technologies | 124 | 21 | 5.9 | 2021 |
| Journal of Business Research | 255 | 10 | 25.5 | 2021 |
| Computers & Education | 234 | 8 | 29.3 | 2020 |
| Information Systems Frontiers | 90 | 7 | 12.9 | 2021 |
| Interactive Learning Environments | 103 | 15 | 6.9 | 2022 |
| Journal of Service Management | 171 | 6 | 28.5 | 2020 |
| Psychology & Marketing | 179 | 9 | 19.9 | 2021 |
| Aesthetic Surgery Journal | 52 | 5 | 10.4 | 2023 |
| British Journal of Educational Technology | 90 | 6 | 15 | 2016 |
| Digital Health | 234 | 9 | 26 | 2019 |
| Educational Technology & Society | 88 | 7 | 12.6 | 2011 |
| Frontiers In Psychology | 111 | 13 | 8.5 | 2019 |
| Frontiers In Public Health | 34 | 5 | 6.8 | 2021 |
| IEEE Access | 112 | 15 | 7.5 | 2019 |
| JMIR Formative Research | 66 | 23 | 2.9 | 2020 |

*NP: Number of publications; TC: Total citations on WoS; AVTC: Average total citation per document; Pub Yr.: Publication Year*

### 4.5.2 Sources Co-Citation Analysis

As part of analyzing the intellectual structure of CGAI application in education and research, we evaluate the sources' co-citation, which helps to address the second research question (RO2) in this study. The co-citation analysis explains the frequency with which pairs of sources (e.g., journals) are cited together. Tracking the co-cited pairs helps to identify the sources that address the same themes in the co-cited network (Egghe and Rousseau, 2002; Donthu et al., 2021; Akpan and Offodile, 2024). It implies that the co-cited journals share some common research themes. As more and more co-citations occur, they form research clusters with higher link strength, making the co-cited source relevant and influential in a research network (Cobo et al., 2011; Akpan and Offodile, 2024).

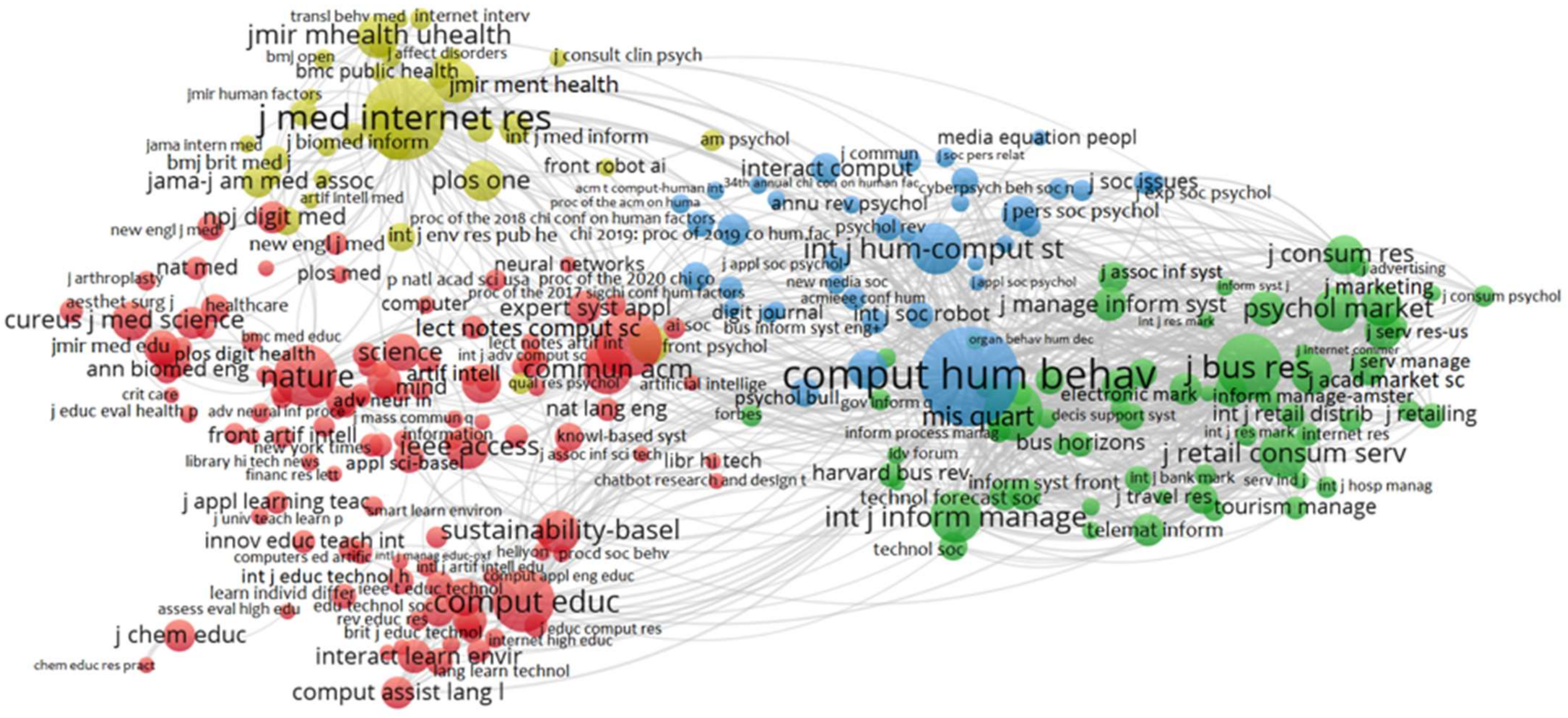

Figure 6. Sources co-citation analysis of research on CGAI

The results summary (Table 2) identified 1025 sources (article: 1424; book chapter: 3; proceedings: 411). The co-citation analysis using the VOSviewer application highlighted the multidisciplinary nature of the sources, which were classified into four (4) color-coded clusters (red, yellow, blue, and green) displayed on a network map (Figure 6). The result shows the interconnectedness among 259 sources that earned at least 40 citations from 2006 to 2023. The leading sources on the co-citation network across the four clusters are:

a. Red Cluster: "Nature," "Communications of the ACM," "Computer Education," and "Lecture Notes in Computer Science;" b. Yellow Cluster: "Journal of Medical Internet Research," "JMIR MHealth and UHealth," "Plos One," and "JAMA;" c. Blue Cluster: "Computers in Human Behavior," "International Journal of Human-Computer Studies," and "Journal of Personal Social Psychology;" d. Green Cluster: "Journal of Business Research," "International Journal of Information Management," and "Journal of Retail Consumer Services." Figure 6 shows the complete list.

These results further reinforce the diverse application domains of CGAI in educational services, research, and other creative activities. Also, the topics and themes highlighted in this study appear in prominent/top quality sources in the various disciplines indicating the importance of AI applications in educational services and research that tends to disrupt the old orders in academics.

## 5. Conclusions and Policy Implications.

The results in this study offer valid evidence that CGAI is undoubtedly affecting teaching, learning, and research and the future of these service operations activities. However, for successful utilization of the CGAI technologies disrupting educational activities, policy guidelines on ethical issues, data privacy and security, intellectual property rights, bias, mitigation, fairness, transparency, and accountability are essential.

Regarding the ethical use considerations, the problems of abuse or misuse of the technology are valid concerns in using ChatGPT for teaching and research (Kooli, 2023; Avenali et al., 2023). Therefore, appropriate guidelines must be developed and followed to establish trust among stakeholders. Also, CGAI (ChatGPT and Chatbot) can generate content written by a human, making it vulnerable to abuse or plagiarism (Heng et al., 2023; Alkaissi and McFarlane, 2023; Adhikari et al., 2023). Learners' data must be protected, and policymakers must clearly outline informed consent sought before use. Policymakers should provide guidelines and appropriate training to avoid misuse.

Regarding data privacy and security, it is essential that organizations develop have strong privacy policies, data encryption, and informed consent processes to protect users' data and maintain trust (Huallpa, et al., 2023). CGAI relies on massive volumes of data to generate responses, including the data used by learners interacting with ChatGPT, Policies must therefore include strict measures to protect student and participant data privacy by providing guidelines for approaches to data collection, storage, and sharing to protect sensitive data in both teaching and research applications. The policy can outline informed consent procedures and handling of data associated with ChatGPT.

"Intellectual property rights" is another area of concern for CGAI use. With the emergence of the ChatGPT platform, it has become necessary that copyright regulations undergo some modifications to tackle the intricacies of authorship and ownership of such AI-generated creative content (Lucchi, 2023). There are issues associated with the use of copyrighted information to train and develop AI systems. The concerns have to do with how such data may be collected and utilized lawfully, as well as the derivative works and fair use. Given the foregoing, the policy needs to provide detailed guidelines to tackle the copyright problems associated with ChatGPT.

The problem of bias and unfairness associated with CGAI in education and research must be addressed. The quality and inclusiveness of the data that ChatGPT received in training determines its effectiveness. When trained on biased data, it will exhibit biases that can result in consequences for individuals or specific groups (Ankita, 2023). To address concerns about biases in AI systems, policies must enforce comprehensive evaluation and mitigation strategies as well as measures to ensure fairness and equity. ChatGPT's performance, particularly in diverse teaching and research settings, must be mandated in the policy document to minimize biases and uphold ethical standards.

Finally, CGAI users must show transparency and accountability. For example, ChatGPT-assisted research must be transparent and accountable, especially when it has to do with research outcomes and implications. This is needed to maintain trust in the academic community and among the general-public (Raman, 2023). The policy must therefore enforce transparency in the use of ChatGPT. Explanations on CGAI integration and details about model architecture, training data, and parameters for educators, students, and other researchers to understand and validate the processes involved. For successful integration of ChatGPT for teaching and research. Policies must specify guidelines to bridge the gap between teachers, researchers, and the public to foster understanding and trust.

## 6. Recommendations and Future Work

The current literature focuses on the need to formulate relevant fair-use policies to contend against the abuse and misuse of GenAI in education and operations activities in different sectors. However, while the policy recommendations (as explained in Section 5) are plausible, legislation can only effectively solve some problems associated with the everyday use of CGAI, implying that educational institutions and organizations can only partially legislate out of the situation. In this Section, we propose new frontiers to tackle the multiple limitations and challenges of GenAI/CGAI.

First and foremost, there is an urgent need to implement and deploy AI-based systems to automatically detect content generated by CGAI/GenAI, which can check potential abuses, such as plagiarism and academic dishonesty. Preferably, such systems need to be discipline-based rather than having a one-size-fits-all approach to ensure effectiveness. For example, a system automatically detecting GenAI natural language text can be less effective in detecting AI-generated computer program codes and mathematical or simulation models used in computer science and operations research for complex systems analysis. The above system can effectively complement GenAI's fair-use policies. Developing such systems can be challenging and hence can be considered an essential further research topic.

Second, in addition to formulating policies to check GenAI abuse and ensure ethical compliance and integrity, educators must carefully consider designing effective learning outcomes and recognizing the new reality. Coursework can be designed so learners can utilize GenAI to complete tasks by borrowing a leaf from the experiment involving a graphing calculator in learning mathematics rather than legislating against its use (Doerr and Zangor, 2000). Curriculum designers and educators can define the roles of GenAI in supporting teaching and instruction in a way that enhances new knowledge and skills development. Such an approach can help rather than hinder innovative education, especially as GenAI use in education and industry appears to have come to stay (Kanbach et al., 2024). Further studies can investigate these possibilities towards breaking new grounds in AI compliance educational and learning design.

Finally, learning and instruction towards knowledge and skills acquisition through educational programs can consider more real-world problems, tasks, or complex scenarios that require critical thinking and creative solutions, encouraging using AI tools as part of the problem-solving toolkit. Evaluation for this endeavor will now focus on course exercises that require analysis, interpretation, and decision-making problems based on data rather than focusing on theoretical knowledge. As discussed in Section 4.5, in the case of OR education, research, and practice, CGAI/GenAI can be implemented to handle some complex situations and enhance idea generation and insights for the learners. Gaining such insights can be instrumental in solving complex problems (Feduzi and Runde, 2014). However, the purpose of integrating GenAI/CGAI in such a learning environment must ensure that GenAI is enhancing rather than replacing the intended skills acquisition, whether it is natural language writing, computer program coding, creating simulation or mathematical model formulation skills (as an application to education and allied industry practice, including OR).